\documentclass[namedreferences]{SolarPhysics}
\usepackage[optionalrh,solaenum]{spr-sola-addons} 
\usepackage{graphicx}
\usepackage{color}
\usepackage{url}             

\newcommand{\aap}{    {\it Astron. Astrophys.}}

\newcommand{\apj}{    {\it Astrophys. J.}}

\newcommand{\jgr}{    {\it J. Geophys. Res.}}

\newcommand{\pasj}{   {\it Pub. Astron. Soc. Japan}}

\newcommand{\solphys}{{\it Solar Phys.}}

\newcommand{\ssr}{    {\it Space Sci. Rev.}}

\begin{document}
\begin{article}
\begin{opening}
\title{Coronal Shock Waves, EUV waves, and their Relation to CMEs.
II. Modeling MHD Shock Wave Propagation Along the Solar Surface,
Using Nonlinear Geometrical Acoustics}

\author{A.N.~\surname{Afanasyev}\sep
        A.M.~\surname{Uralov}}

\runningauthor{Afanasyev and Uralov}

\runningtitle{Modeling MHD shock wave propagation}

\institute{Institute of Solar-Terrestrial Physics SB RAS,\\P.O. Box
291, Lermontov~St.~126A, Irkutsk 664033, Russia
email:~\url{afa@iszf.irk.ru} email:~\url{uralov@iszf.irk.ru}}

\begin{abstract}
We model the propagation of a coronal shock wave, using nonlinear
geometrical acoustics. The method is based on the
Wentzel--Kramers--Brillouin (WKB) approach and takes into account
the main properties of nonlinear waves: \textit{i})~dependence of
the wave front velocity on the wave amplitude,
\textit{ii})~nonlinear dissipation of the wave energy, and
\textit{iii})~progressive increase in the duration of solitary
shock waves. We address the method in detail and present results
of the modeling of the propagation of shock-associated
extreme-ultraviolet (EUV) waves as well as Moreton waves along the
solar surface in the simplest solar corona model. The calculations
reveal deceleration and lengthening of the waves. In contrast,
waves considered in the linear approximation keep their length
unchanged and slightly accelerate.
\end{abstract}

\end{opening}

\section{Introduction}

Many solar eruptive events appear to initiate magnetohydrodynamic
(MHD) waves in the corona. This conjecture is supported by
observations of their manifestations. First, these are waves visible
in chromospheric spectral lines. For several decades, Moreton waves
have been known  \cite{Moreton60} observed in the H$\alpha$ line.
\inlinecite{Uchida68} proposed that a Moreton wave represented the
chromospheric trail of a coronal fast-mode wave. More recent studies
show the possible coronal nature of Moreton waves (see,
\textit{e.g.}, \opencite{Balasubramaniam07}). Similar phenomena are
propagating waves observed in the He I 10830 $\mathrm \AA$ line
(see, \textit{e.g.}, \opencite{Vrsnak02b}; \opencite{Gilbert04}).
Second, type II radio bursts are considered to be signatures of
coronal shock waves propagating upwards in the corona
\cite{Uchida60}. Furthermore, presumable signatures of coronal waves
can be observed in soft X-rays (\opencite{Narukage02};
\opencite{Khan02}; \opencite{Hudson03}; \opencite{Warmuth05}),
microwaves (\opencite{Warmuth04a}; \opencite{White05}), and metric
radio-wavelengths (\opencite{Vrsnak05}).

Other phenomena that can be manifestations of MHD waves are
large-scale wave-like disturbances discovered in 1998 with the
Extreme-ultraviolet Imaging Telescope (EIT) and referred to as
``EIT waves'' or ``extreme-ultraviolet (EUV) waves''. This term
appears to include phenomena having different physical nature and
therefore various morphological and dynamical properties. For more
details, we refer the reader to the Introduction of an
accompanying paper \cite{Grechnev10I}. Note here that describing
EUV waves in terms of fast-mode MHD waves seems to be possible and
correct for those cases when we really deal with ``wave''
phenomena. Just this class of EUV transients is the subject of our
consideration.

Thus, fast-mode MHD waves are responsible for a number of solar
transients and it is important to describe their propagation
through the corona. But some questions about modeling the waves
remain.

In his original approach, \inlinecite{Uchida68} modeled a coronal
disturbance as a linear and short fast-mode MHD wave propagating
from a point source. To calculate a dome-like surface of the wave
front and find the position of a Moreton wave (as an intersection
line of the dome and the solar surface), he used the linear
geometric acoustics [the Wentzel--Kramers--Brillouin (WKB)
approach]. Some recent studies (see, \textit{e.g.},
\opencite{Wang00}; \opencite{Patsourakos09}) also used the same
approach under the assumption of a linear disturbance. In this case,
the disturbance moves along rays, which curve into regions of a
reduced Alfv{\'e}n speed. This results in the appearance of wave
``imprints'' (\textit{e.g.}, Moreton waves) running along the
spherical solar surface. Note that in the linear approximation the
amplitude and duration of the wave do not affect the shape of the
wave front and the speed of its motion along the rays. Neither
amplitude nor duration of the wave were calculated in the mentioned
papers; the geometry of the propagation was their only interest.
Henceforth in the paper, we consider the wave amplitude to be the
perturbation magnitude of the plasma velocity in the wave.

Uchida's model of a linear MHD wave has demonstrated the possibility
to describe Moreton waves in these terms. Later papers by Uchida
with collaborators (\opencite{Uchida73}; \opencite{Uchida74}) and
recent papers (\opencite{Wang00}; \opencite{Patsourakos09}), in
which the coronal magnetic field was calculated from photospheric
magnetograms, provided a more accurate quantitative description of
Moreton and EUV waves in terms of the linear model.

However, the linear model predicts acceleration of Moreton and EUV
waves, whereas observations show their systematic deceleration
(\citeauthor{Warmuth01} \citeyear{Warmuth01};
\citeyear{Warmuth04a}). Also, it is often pointed out by many
authors that the speeds of coronal waves sometimes well exceed the
fast-mode ones (\opencite{Narukage02}; \opencite{Vrsnak02a};
\opencite{Narukage04}; \opencite{Warmuth04b}; \opencite{Muhr10}).
These facts suggest that the linear approximation does not always
correctly describe the propagation of those waves. Probably, the
disturbance responsible for the transients listed above is nonlinear
and most likely is a shock wave.

Studies of the propagation of shock waves meet difficulties due to
their nonlinearity. Analytic methods to describe the propagation of
shock waves are approximate and often describe the behavior of some
extreme classes of nonlinear waves (\textit{e.g.}, very strong
self-similar waves, or weak waves, \textit{etc.}). The present paper
is devoted to the case of a weak shock wave that appears to be the
most acceptable (see, \textit{e.g.}, \opencite{Warmuth04b};
\opencite{Vrsnak02a}). We do not discuss the appearance of a shock
wave. \inlinecite{Vrsnak00} approximately described this process,
based on an analogy with an accelerating flat piston. The cases of
cylindrical and spherical pistons were analyzed by
\inlinecite{Zic08}. \inlinecite{Temmer09} demonstrated how one could
describe the kinematics of a Moreton wave by using the solution of a
simple wave without a discontinuity. In our study, we assume that a
fast-mode shock wave of moderate intensity appears during a solar
eruption on the periphery of an active region and decays to a weak
shock when traveling in the corona. The wave manifests as a Moreton
wave and an EUV wave on the solar disk. We calculate the propagation
of the shock wave in terms of the WKB approach taking into account
nonlinear effects.

Such a method in its generally accepted variant involves two
independent procedures. In the first one, ray trajectories
corresponding to the linear approximation as well as cross sections
of ray tubes are calculated. So, the influence of the finite wave
amplitude on the wave front shape and ray trajectories is ignored
because in linear acoustics the propagation velocities of
disturbances are equal to the undisturbed sound speed regardless of
their amplitude. In the second procedure, the nonlinear variation of
the wave amplitude and duration are computed along the linear rays
obtained. Such a non-self-consistent approach is fairly useful in
some cases, however, nonlinear effects altering the ray pattern and
the wave velocity disappear completely. Therefore, we develop
another method, which allows us to consider self-consistently wave
propagation and the nonlinear variation of wave characteristics.

The method is described in Section~\ref{S-method}. In
Section~\ref{S-modeling} we formulate the problem and present
results of the analytic propagation model of Moreton and EUV waves
along the solar surface. Section~\ref{S-conclusion} contains
concluding remarks about the method and the results. We note that
those who are not interested in the mathematical details of the
method can read Section 3 without going through Section~2.

\section{Method}
\label{S-method}

The method of nonlinear geometrical acoustics is based on the method
of linear geometrical acoustics and allows one to calculate the
propagation of disturbances with small (but finite) amplitudes
through an inhomogeneous medium. The linear geometrical acoustics is
known to be a method to calculate linear disturbances in the ray
approximation (see, \textit{e.g.}, \opencite{Landau87}). In this
approximation, a solution is found in the form of $A\left(
\mathbf{r},t\right) e^{i\Psi \left( \mathbf{r},t\right) }$ where
$A\left( \mathbf{r},t\right)$ is the wave amplitude, and $\Psi
\left( \mathbf{r},t\right)$ is the eikonal, both depending on
coordinates and time. By substituting this representation for wave
perturbations into the system of linearized equations of ideal
magnetohydrodynamics, one can obtain a Hamilton--Jacobi partial
differential equation for the eikonal of fast and slow magnetosonic
waves:
$$\frac{\partial \Psi }{\partial
t}+\left( \mathbf{V}\mathrm{grad}\Psi \right) +a\left\vert
\mathrm{grad}\Psi \right\vert =0,$$ where $\mathbf{V}$ is the
undisturbed plasma  flow velocity (\textit{e.g.}, the solar wind
velocity), and $a$ is the magnetosonic speed in plasma. Solving the
equation with the method of characteristics gives a system of ray
equations, which in spherical coordinates $(r,\theta, \varphi)$
takes the form \cite{Uralova94}
%
\begin{eqnarray}
 \frac{dr}{dt}&=&V_{r}+a\frac{k_{r}}{k}+k\frac{\partial a}{\partial k_{r}},
 \nonumber \\
 r\frac{d\theta }{dt}&=&a\frac{k_{\theta }}{k}+k\frac{\partial a}{\partial
 k_{\theta }}, \nonumber  \\
 r\sin \theta \frac{d\varphi }{dt}&=&a\frac{k_{\varphi }}{k}+k\frac{\partial
 a}{\partial k_{\varphi }},
 \label{E-ray_eq_system} \\
 \frac{dk_{r}}{dt}&=&-\frac{\partial V_{r}}{\partial r}k_{r}-
 \frac{\partial a}{\partial r}k+\frac{a}{kr}\left( k_{\theta }^{2}+
 k_{\varphi}^{2}\right), \nonumber \\
 r\frac{dk_{\theta }}{dt}&=&-\frac{\partial V_{r}}{\partial \theta }k_{r}-
  \frac{\partial a}{\partial \theta }k+\frac{a}{k}k_{\varphi}^{2}\cot \theta-
 k_{\theta }\frac{dr}{dt},  \nonumber \\
 r\sin \theta \frac{dk_{\varphi }}{dt}&=&-\frac{\partial V_{r}}{\partial
 \varphi }k_{r}-\frac{\partial a}{\partial \varphi }k-\sin \theta
 k_{\varphi } \frac{dr}{dt}-k_{\varphi }r\cos \theta \frac{d\theta}{dt},
 \nonumber
\end{eqnarray}
where $k_{r,\theta,\varphi}$ are the components of the wave vector
$\mathbf{k} = \mathrm{grad} \Psi$, and $k$ is its magnitude. The
system of equations~(\ref{E-ray_eq_system}) corresponds to the case
of a medium in steady-state, where only the radial component $V_r$
of the undisturbed plasma flow exists. By integrating the
system~(\ref{E-ray_eq_system}), one can determine the wave front
shape.

This approach has been used for modeling coronal fast-mode MHD waves
\cite{Uchida68,Wang00}, with only the ray pattern being calculated.
However, it is essential not only to find the wave geometry, but
also to calculate the wave intensity. The geometrical acoustics
allows the wave amplitude variation to be calculated.

In the linear geometrical acoustics, the energy flux of a
disturbance traveling in a stationary medium with group velocity
$\mathbf{q}_0$ is directed along the rays, and its magnitude is
conserved within a ray tube \cite{Blokhintsev81},
$\mathrm{div}(\Delta \varepsilon \mathbf{q}_0) = 0$, where $\Delta
\varepsilon$ is the average density of the disturbance energy. In a
medium moving at a velocity of $\mathbf{V}$, we have to take into
account the fact that the wave front phase velocity varies as $q_n =
V_n+a$, with the $n$ index denoting the projection normal to the
front. The conservation law in this case is $\mathrm{div} (\Delta
\varepsilon \mathbf{q}\, q_{n}/a) = 0$ \cite{Uralov82,Barnes92}
where $\mathbf{q} = \mathbf{V} + \mathbf{q}_0$ is the group velocity
in a moving medium. The average density of the disturbance energy is
$\Delta \varepsilon =\rho \overline{\left( u^{2}+v^{2}\right) }$
where $\rho$ is the undisturbed plasma density, and $u,v$ are the
plasma velocity components along the normal to the wave front and
across it, respectively. Taking into account the relation between
the plasma velocity components $\mu = v/u$ \cite{Kulikovsky05}, it
is possible to relate the variation of the wave amplitude to the
normal cross section $dS$ of the ray tube formed by a bundle of
rays:
\begin{equation}
 dSq\rho u^{2}\left( 1+\mu ^{2}\right) \frac{q_{n}}{a}=const.
 \label{E-dS_wave_amp_variation}
\end{equation}
Thus, the general approach to determine the wave amplitude in the
ray method relies on calculating the cross section of the ray tube.

There are various techniques for calculating cross sections. They
are discussed in \inlinecite{KravtsovOrlov90}. We calculate cross
sections by using the Jacobian of the transformation to ray
coordinates. The volume element $dW$ of the ray tube is expressed in
terms of ray coordinates $(\eta _{1}, \eta _{2}, t)$ as
$$dW=dxdydz=r^{2}\sin \theta drd\theta d\varphi =r^{2}\sin \theta D\left(
t\right) d\eta _{1}d\eta _{2}dt,$$ where $D(t)$ is the Jacobian of
the transformation from spherical coordinates to ray ones. Then for
the cross section of the ray tube, $dS$, we have
$$dS=\frac{dW}{d\sigma }=r^{2}\sin \theta \frac{D\left( t\right)
}{q}d\eta _{1}d\eta _{2},$$ where $\sigma$ is the ray tube length.
Substituting it into (\ref{E-dS_wave_amp_variation}) yields
\begin{equation}
D\left( t\right) r^{2}\sin \theta \rho u^{2}\left( 1+\mu ^{2}\right) \frac{%
q_{n}}{a}=const. \label{E-Jacob_wave_amp_variation}
\end{equation}

To calculate the Jacobian, we use a method based on numerical
integration of a so-called adjoint system of equations. This system
consists of differential equations for the derivatives $\partial
r/\partial \eta _{1,2}$, $ \partial \theta/\partial \eta _{1,2}$, $
\partial \varphi/\partial \eta _{1,2}$, $\partial
k_r/\partial \eta _{1,2}$, $ \partial k_{\theta}/\partial \eta
_{1,2}$, $
\partial k_{\varphi}/\partial \eta _{1,2}$  and is derived from
(\ref{E-ray_eq_system}) by differentiating the equations with
respect to ray coordinates $\eta _{1}$ and $\eta _{2}$. Let the ray
coordinates $\eta _{1}$ and $\eta _{2}$ be the angles defining the
direction of the outgoing ray from a point source at the initial
moment. Note that the ray coordinates do not need to be explicitly
determined when the adjoint system is being derived. This becomes
essential to specify the initial values for the desired functions.
For the case considered, the adjoint system has the following form
(in view of the symmetry about $\eta _{1}$ and $\eta _{2}$, instead
of 12 equations we give only six for one variable $\eta$):
%
\begin{eqnarray}
\frac{d}{dt}\left( \frac{\partial r}{\partial \eta }\right)
=\frac{\partial
V_{r}}{\partial \eta }+\frac{k_{r}}{k}\frac{\partial a}{\partial \eta }+%
\frac{a}{k}\frac{\partial k_{r}}{\partial \eta }-\frac{ak_{r}}{k^{2}}\frac{%
\partial k}{\partial \eta }+\frac{\partial a}{\partial k_{r}}\frac{\partial k%
}{\partial \eta }+k\frac{\partial }{\partial \eta }\left(
\frac{\partial
a}{\partial k_{r}}\right), \nonumber \\
\frac{\partial r}{\partial \eta }\frac{d\theta
}{dt}+r\frac{d}{dt}\left(
\frac{\partial \theta }{\partial \eta }\right) =\frac{k_{\theta }}{k}\frac{%
\partial a}{\partial \eta }+\frac{a}{k}\frac{\partial k_{\theta }}{\partial
\eta }-\frac{ak_{\theta }}{k^{2}}\frac{\partial k}{\partial \eta }+\frac{%
\partial a}{\partial k_{\theta }}\frac{\partial k}{\partial \eta }+k\frac{%
\partial }{\partial \eta }\left( \frac{\partial a}{\partial k_{\theta }}%
\right), \nonumber \\
\frac{\partial r}{\partial \eta }\sin \theta \frac{d\varphi
}{dt}+r\cos \theta \frac{\partial \theta }{\partial \eta
}\frac{d\varphi }{dt}+r\sin \theta \frac{d}{dt}\left(
\frac{\partial \varphi }{\partial \eta }\right) =\qquad
\qquad\qquad\qquad\qquad\qquad \nonumber \\
=\frac{\partial a}{\partial \eta }\frac{k_{\varphi }}{k}+\frac{a}{k}\frac{\partial k_{\varphi }}{\partial \eta }-%
\frac{ak_{\varphi }}{k^{2}}\frac{\partial k}{\partial \eta }+\frac{\partial a%
}{\partial k_{\varphi }}\frac{\partial k}{\partial \eta }+k\frac{\partial }{%
\partial \eta }\left( \frac{\partial a}{\partial k_{\varphi
}}\right) , \nonumber \\
\frac{d}{dt}\left( \frac{\partial k_{r}}{\partial \eta }\right)%
=-\frac{\partial V_{r}}{\partial r}\frac{\partial k_{r}}{\partial \eta }-k_{r}\frac{%
\partial }{\partial \eta }\left( \frac{\partial V_{r}}{\partial r}\right) -%
\frac{\partial a}{\partial r}\frac{\partial k}{\partial \eta }-k\frac{%
\partial }{\partial \eta }\left( \frac{\partial a}{\partial r}\right)+
\frac{\partial a}{\partial \eta }\frac{k_{\theta }^{2}+k_{\varphi
}^{2}}{kr}- \nonumber \\
-\frac{a}{k^{2}r}\left( k_{\theta }^{2}+k_{\varphi }^{2}\right)
\frac{\partial k}{\partial \eta }-\frac{a\left( k_{\theta
}^{2}+k_{\varphi }^{2}\right) }{kr^{2}}\frac{\partial r}{\partial \eta }+
\frac{a}{kr}\left( 2k_{\theta }\frac{\partial k_{\theta }}{\partial \eta }%
+2k_{\varphi }\frac{\partial k_{\varphi }}{\partial \eta }\right) , \nonumber \\
\frac{\partial r}{\partial \eta }\frac{dk_{\theta }}{dt}+r\frac{d}{dt}%
\left( \frac{\partial k_{\theta }}{\partial \eta }\right)
=-\frac{\partial
V_{r}}{\partial \theta }\frac{\partial k_{r}}{\partial \eta }-k_{r}\frac{%
\partial }{\partial \eta }\left( \frac{\partial V_{r}}{\partial \theta }%
\right) -\frac{\partial a}{\partial \theta }\frac{\partial k}{\partial \eta }%
-k\frac{\partial }{\partial \eta }\left( \frac{\partial a}{\partial \theta }%
\right)+ \nonumber \\
+\frac{\partial a}{\partial \eta }\frac{k_{\varphi
}^{2}\cot \theta }{k}-\frac{ak_{\varphi }^{2}}{k^{2}}\cot \theta \frac{\partial k%
}{\partial \eta }+\frac{2ak_{\varphi }}{k}\cot \theta \frac{\partial
k_{\varphi }}{\partial \eta }-\qquad\qquad\qquad\qquad \nonumber \\
-\frac{ak_{\varphi }^{2}}{k\sin ^{2}\theta }\frac{\partial \theta
}{\partial
\eta }-\frac{\partial k_{\theta }}{\partial \eta }\frac{dr}{dt}-k_{\theta }%
\frac{d}{dt}\left(
\frac{\partial r}{\partial \eta }\right),\label{E-eq_system_for_calc amp}  \\
\frac{\partial r}{\partial \eta }\sin \theta \frac{dk_{\varphi
}}{dt}+r\cos \theta \frac{\partial \theta }{\partial \eta
}\frac{dk_{\varphi }}{dt}+r\sin
\theta \frac{d}{dt}\left( \frac{\partial k_{\varphi }}{\partial \eta }%
\right) = \qquad\qquad\qquad\qquad\qquad\qquad \nonumber \\
=-\frac{\partial V_{r}}{\partial \varphi }\frac{\partial k_{r}}{%
\partial \eta }-k_{r}\frac{\partial }{\partial \eta }\left( \frac{\partial
V_{r}}{\partial \varphi }\right) -\frac{\partial a}{\partial \varphi }\frac{%
\partial k}{\partial \eta }-k\frac{\partial }{\partial \eta }\left( \frac{%
\partial a}{\partial \varphi }\right) -\cos \theta \frac{\partial \theta }{%
\partial \eta }k_{\varphi }\frac{dr}{dt}- \nonumber \\
-\sin \theta \frac{\partial k_{\varphi }}{\partial \eta }%
\frac{dr}{dt}-\sin \theta k_{\varphi }\frac{d}{dt}\left( \frac{\partial r}{%
\partial \eta }\right)-\frac{\partial k_{\varphi }}{\partial \eta }r\cos \theta
\frac{d\theta }{dt}-k_{\varphi }\frac{\partial r}{\partial \eta
}\cos \theta \frac{d\theta }{dt}+ \nonumber \\
+k_{\varphi
}r\sin \theta \frac{\partial \theta }{\partial \eta }\frac{d\theta }{dt}%
-k_{\varphi }r\cos \theta \frac{d}{dt}\left( \frac{\partial \theta }{%
\partial \eta }\right) , \nonumber
\end{eqnarray}
where
\begin{eqnarray*}
\frac{\partial V_{r}}{\partial \eta }=\sum\limits_{\alpha
}\frac{\partial V_{r}}{\partial r_{\alpha }}\frac{\partial r_{\alpha
}}{\partial \eta },
\quad\frac{\partial a}{\partial \eta
}=\sum\limits_{\alpha }\left( \frac{\partial a}{\partial r_{\alpha
}}\frac{\partial r_{\alpha }}{\partial \eta }+\frac{
\partial a}{\partial k_{\alpha }}\frac{\partial k_{\alpha }}{\partial \eta }
\right),
\quad\frac{\partial k}{\partial
\eta}=\frac{1}{k}\sum\limits_{\alpha }k_{\alpha }\frac{\partial
k_{\alpha }}
{\partial \eta}, \\
\frac{\partial }{\partial \eta }\left( \frac{\partial a}{\partial k_{\beta }%
}\right) =\sum\limits_{\alpha }\left( \frac{\partial }{\partial r_{\alpha }}%
\left( \frac{\partial a}{\partial k_{\beta }}\right) \frac{\partial
r_{\alpha }}{\partial \eta }+\frac{\partial }{\partial k_{\alpha
}}\left(
\frac{\partial a}{\partial k_{\beta }}\right) \frac{\partial k_{\alpha }}{%
\partial \eta }\right),\qquad\qquad \\
\frac{\partial }{\partial \eta }\left( \frac{\partial a}{\partial r_{\beta }%
}\right) =\sum\limits_{\alpha }\left( \frac{\partial }{\partial r_{\alpha }}%
\left( \frac{\partial a}{\partial r_{\beta }}\right) \frac{\partial
r_{\alpha }}{\partial \eta }+\frac{\partial }{\partial k_{\alpha
}}\left(
\frac{\partial a}{\partial r_{\beta }}\right) \frac{\partial k_{\alpha }}{%
\partial \eta }\right), \qquad\qquad \\
\frac{\partial }{\partial \eta }\left( \frac{%
\partial V_{r}}{\partial r_{\beta }}\right) =\sum\limits_{\alpha }\left(
\frac{\partial }{\partial r_{\alpha }}\left( \frac{\partial
V_{r}}{\partial r_{\beta }}\right) \frac{\partial r_{\alpha
}}{\partial \eta }\right), \quad r_{\alpha,\beta}=\{r,\theta
,\varphi \}, \; \; k_{\alpha,\beta}=\{k_{r},k_{\theta },k_{\varphi
}\}.
\end{eqnarray*}

Thus, to calculate the propagation of a linear wave and its
intensity, at first we have to integrate numerically the ray
equations system~(\ref{E-ray_eq_system}) and adjoint
system~(\ref{E-eq_system_for_calc amp}) and then determine the
amplitude variations by means
of~(\ref{E-Jacob_wave_amp_variation}).

A nonlinear flat disturbance in an ideal homogeneous medium is
described by a simple wave solution and propagates at the supersonic
speed determined by its amplitude \cite{Kulikovsky05}. A fast-mode
simple-wave element with plasma velocity component $u$ normal to the
front moves at $a+\kappa u$, where $\kappa =\left( 1/a\right) \left(
d\left( \rho a\right) /d\rho \right) $ is the numerical coefficient
depending both on the plasma beta and the angle between the wave
vector and the magnetic field. We do not give here the bulky
explicit expression for $\kappa$. We only note that values of
$\kappa$ are restricted by the limits: $1/2\leq \kappa \leq 3/2$.
The fact that each simple-wave element travels at its own speed
causes wave profile deformation and the appearance of a
discontinuity. If a moderate amplitude simple wave has a triangular
profile before the discontinuity appears, it will take the shape of
the right-angled triangle after the discontinuity forms, with the
discontinuity being the leading edge of the disturbance. Note that
any nonlinear disturbance profile of a finite duration tends
asymptotically to this shape. Let $U_{sh}$ be the jump of the plasma
velocity component $u$ in the discontinuity. Then, in the nonlinear
geometrical acoustics approximation, the discontinuity moves at a
speed of $a+\kappa U_{sh}/2$. Taking into account this increase in
the wave front speed in the ray equations, we are able to correctly
describe the propagation of weak shock waves. Then the ray equation
system~(\ref{E-ray_eq_system}) becomes \cite{Uralova94}
%
\begin{eqnarray}
 \frac{dr}{dt}&=&V_{r}+\left( a+\frac{\kappa U_{sh}}{2}\right) \frac{k_{r}}{k}%
 +k\frac{\partial a}{\partial k_{r}}, \nonumber \\
 r\frac{d\theta }{dt}&=&\left( a+\frac{\kappa U_{sh}}{2}\right) \frac{%
 k_{\theta }}{k}+k\frac{\partial a}{\partial k_{\theta }}, \nonumber \\
 r\sin \theta \frac{d\varphi }{dt}&=&\left( a+\frac{\kappa
 U_{sh}}{2}\right) \frac{k_{\varphi }}{k}+k\frac{\partial a}{\partial
 k_{\varphi }},\label{E-nonlin_ray_eq_system} \\
 \frac{dk_{r}}{dt}&=&-\frac{\partial V_{r}}{\partial r}k_{r}-\frac{\partial a}{%
 \partial r}k+\frac{a}{kr}\left( k_{\theta }^{2}+k_{\varphi }^{2}\right)
 , \nonumber \\
 r\frac{dk_{\theta }}{dt}&=&-\frac{\partial V_{r}}{\partial \theta }k_{r}-%
 \frac{\partial a}{\partial \theta }k+\frac{a}{k}k_{\varphi
 }^{2}\cot\theta -k_{\theta }\frac{dr}{dt}, \nonumber \\
 r\sin \theta \frac{dk_{\varphi }}{dt}&=&-\frac{\partial
 V_{r}}{\partial
 \varphi }k_{r}-\frac{\partial a}{\partial \varphi }k-\sin \theta k_{\varphi }%
 \frac{dr}{dt}-k_{\varphi }r\cos \theta \frac{d\theta }{dt}. \nonumber
\end{eqnarray}

The generally accepted method of nonlinear geometrical acoustics
ignores the additional term $\kappa U_{sh}/2$ as it is small.
However, it is the term that is responsible for wave deceleration
due to the amplitude damping. Besides, estimations within the
framework of the perturbation theory suggest that the ray pattern
variation due to nonlinearity is a correction of the same order of
magnitude as the nonlinear variation of the wave amplitude is. It is
therefore important to take this into account in the nonlinear
geometrical acoustics approximation.

Ray equation system~(\ref{E-nonlin_ray_eq_system}) is not closed now
because it includes the wave amplitude. In the linear approximation,
an amplitude variation can be determined
from~(\ref{E-Jacob_wave_amp_variation}). The nonlinear wave
amplitude undergoes additional damping associated with energy
dissipation in the discontinuity. As the amplitude, we take the
value of the jump $U_{sh}$. Variations of amplitude $U_{sh}$ and
duration $T_{sh}$ of a weak shock wave having a triangular
compression phase may be calculated as \cite{Uralov82}
\begin{equation}
U_{sh}=u_{1}\left( 1+\frac{\tau _{1}}{T_{\ast }}\right)^{-1/2}, \;\;
T_{sh}=T_{\ast }\left( 1+\frac{\tau _{1}}{T_{\ast }}\right) ^{1/2},
\;\; \frac{d\tau _{1}}{dt}=\frac{\kappa
u_{1}}{q_n}\label{E-damping_laws}
\end{equation}
where $\tau_{1}$ is the duration increment of the simple wave with
an amplitude of $u_{1}$; $T_{\ast}$ is the initial duration of the
disturbance. Note that the laws (\ref{E-damping_laws}) of the weak
shock wave damping are derived by using values of the amplitude and
duration of a simple wave, from which the discontinuity forms. The
value of $u_{1}$ can be determined from the expression similar
to~(\ref{E-Jacob_wave_amp_variation}).

Thus, solving numerically of 19 ordinary differential equations
(\ref{E-nonlin_ray_eq_system}), (\ref{E-eq_system_for_calc amp}),
and (\ref{E-damping_laws}) enables us to compute propagation of a
weak shock wave in an inhomogeneous medium, its amplitude and
duration.

\section{Analytical Modeling of Wave Propagation}
\label{S-modeling}

In this section, we employ the nonlinear geometrical acoustics
method to describe the propagation of large-scale wave-like
transients, namely EUV and Moreton waves. With respect to ``EUV
wave'' phenomena, we address only those disturbances that are
associated with a fast-mode MHD shock wave.

To calculate the propagation of a shock wave, we have to specify the
solar corona model as well as position and parameters of the shock
wave at the initial moment. We use a simple hydrostatic corona model
to demonstrate the main particularities of the method and compare
results with those obtained in the linear approximation. The corona
is considered to be isothermal with temperature $T=1.5\times 10^6$~K
and sound speed $c=181$ km s$^{-1}$. The plasma density is
distributed in accordance with the barometric law (for details see,
\textit{e.g.}, \opencite{Mann99}):
\begin{equation}
n(r)=n_0 \;\mathrm{exp} \left( \frac{R_\odot}{H} \left(
\frac{R_\odot}{r}-1\right) \right),\label{E-barometric_density_law}
\end{equation}
where $n_0=n(R_\odot)=3\times 10^8$ cm$^{-3}$ is the density at the
base of the corona, $R_\odot$ the solar radius,
$H=2R_{gas}T/\widetilde m M_H g_\odot\approx70$ Mm the density scale
height, $g_\odot$ the acceleration of gravity on the solar surface,
$M_H$ the molar mass of hydrogen, $\widetilde{m}=1.27$ the average
atomic weight of an ion, and $R_{gas}$ the gas constant. Let us
assume that we have a magnetic field with only a radial component
(for details see, \textit{e.g.}, \opencite{Mann03}):
\begin{equation}
B_r=\pm B_0 \left( \frac{R_\odot}{r} \right)^2,\label{E-magn_field}
\end{equation}
where $B_0=2.3$ G is its value at the base of the corona. The sign
in (\ref{E-magn_field}) depends on the solar hemisphere, but it is
not meaningful here. The same model was applied by
\inlinecite{Uchida68}.

\begin{figure} 
  \centerline{\includegraphics[width=0.5\textwidth]{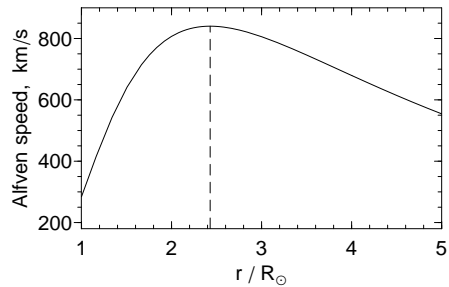} }
\caption{The Alfv{\'e}n speed distribution in the solar corona
model (\ref{E-barometric_density_law}), (\ref{E-magn_field}).}
  \label{F-alfven_velocity}
 \end{figure}

In the corona model (\ref{E-barometric_density_law}) and
(\ref{E-magn_field}), the Alfv{\'e}n speed increases with height,
peaking at $R_\odot^2/4H=2.43\;R_\odot$
(Figure~\ref{F-alfven_velocity}). Refraction makes ray trajectories
curved towards regions of the lower Alfv{\'e}n speed. The solar
corona has therefore waveguide properties. A portion of the wave
energy flux is captured by the coronal waveguide and propagates
along the solar surface, giving rise to an EUV and Moreton wave. In
the treatment used, the EUV front is observable due to the plasma
compression produced by the coronal shock wave. Since the plasma
density decreases rapidly with height, a plasma layer near the solar
surface contributes substantially to the EUV front emission. The
layer thickness is about the density scale height $H$. So, to
estimate the EUV front position, we have to find the intersection
line of the calculated shock front and a spherical surface of radius
$\lesssim (R_\odot +H)$. The Moreton wave corresponds to the
chromospheric trail of the coronal shock wave, \textit {i.e.} it can
be found at the intersection line of the shock front and the upper
chromosphere.

\begin{figure} 
  \centerline{\includegraphics[width=\textwidth]{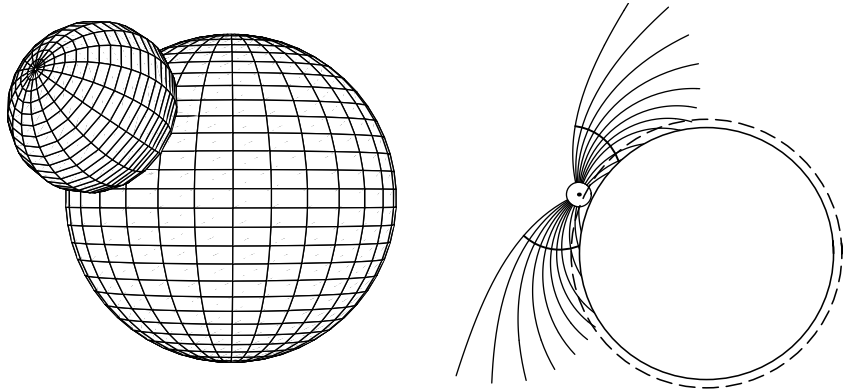} }
\caption{Propagation of a coronal shock wave as seen in a 3D image
(left) and a 2D cross section with ray trajectories (right). The
solid arcs drawn across the rays represent shock wave fronts. The
dashed line marks a height of 60~Mm above the solar surface. The
wave front velocity along this line corresponds to the velocity of
an EUV wave. The wave source height is 80 Mm.}
  \label{F-wave_prop}
 \end{figure}

\begin{figure} 
  \centerline{\includegraphics[width=0.5\textwidth]{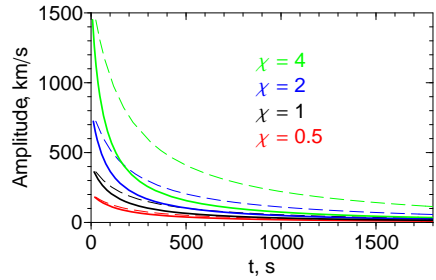} }
\caption{The amplitude damping of shock waves (solid lines) and
linear ones (dashed lines) calculated for the wave source energy
$\tilde{\varepsilon} = 10^{29}$~erg and different initial
amplitudes. The initial amplitude values are specified in the figure
as ratios $\chi$ of those to a fast-mode speed in the wave source.}
  \label{F-EUV_decay}
 \end{figure}

For modeling, it is important to assign the initial characteristics
of the shock wave and its start position. We have to specify the
initial duration (or length) of the wave and its initial amplitude
on some surface. In this study, these values are given according to
the strong point-like explosion theory. The wave source located at a
height of 80 Mm is characterized by the energy $\tilde \varepsilon $
whose release produces the shock wave. When the wave covers a
distance of $\Lambda=(\tilde\varepsilon/\rho_\ast a_\ast^2)^{1/3}$,
with $\rho_\ast$ and $a_\ast$ being respectively the plasma density
and the fast-mode speed at the explosion point, the compression
phase profile of the wave is assumed to be triangular. The
compression phase length is equal to $\Lambda$ and the amplitude is
$\chi a_\ast$, with $\chi$ being a coefficient of the order of
unity. In this paper, we employ a value $\chi=1$ except for the
calculations given in Figure~\ref{F-EUV_decay}. The initial duration
of the compression phase is supposed to be equal to
$T_\ast=\Lambda(\mathbf{k})/a_\ast(\mathbf{k})$. We believe that a
shock wave arises on the periphery of an active region located
within the explosion cavity $\Lambda$. This manner to assign initial
values does not rely on the specific mechanism of the wave
initiation. It is essential only that the energy release producing a
shock wave is impulsive. For instance, a wave can be produced by a
compact piston acting for a short term (an abruptly accelerating
filament and its magnetic envelope).

\begin{figure} 
  \centerline{  \includegraphics[width=0.5\textwidth,clip=]{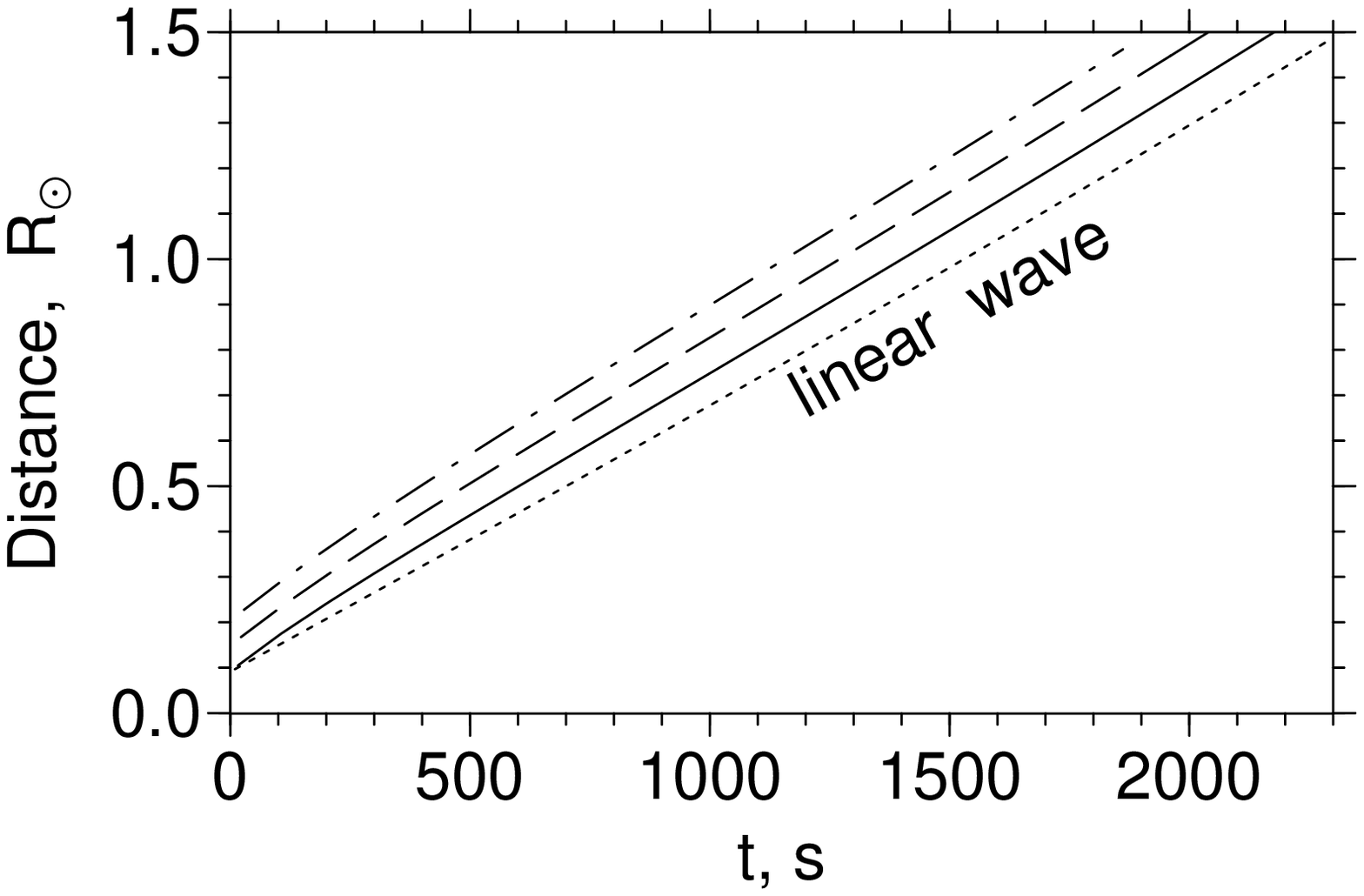}
                \includegraphics[width=0.5\textwidth,clip=]{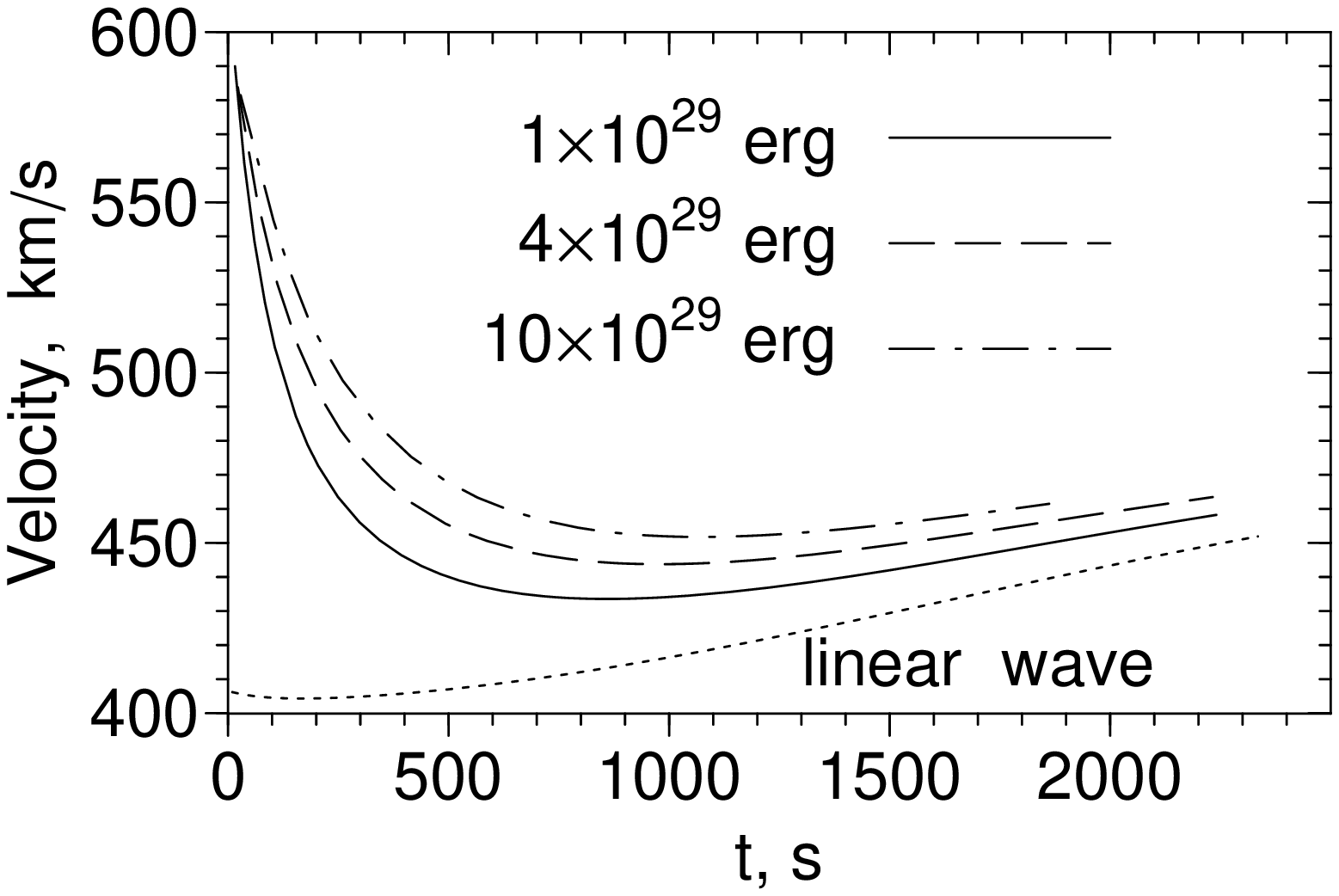}
             }
    \vspace{-0.11\textwidth}
          \centerline{\large
      \hspace{0.425 \textwidth}  {a}
      \hspace{0.47 \textwidth} {b}
         \hfill}
   \vspace{0.075\textwidth}

\caption{The distance--time plot of EUV waves (a) and the time
evolution of their velocities (b) along the surface at 60~Mm height
for different energies $\tilde{\varepsilon}$ specified in panel b.
The lowest (dotted) line represents the linear EUV wave.}
  \label{F-wave_speed_vs_energy}
  \end{figure}

\begin{figure} 
  \centerline{  \includegraphics[width=0.5\textwidth,clip=]{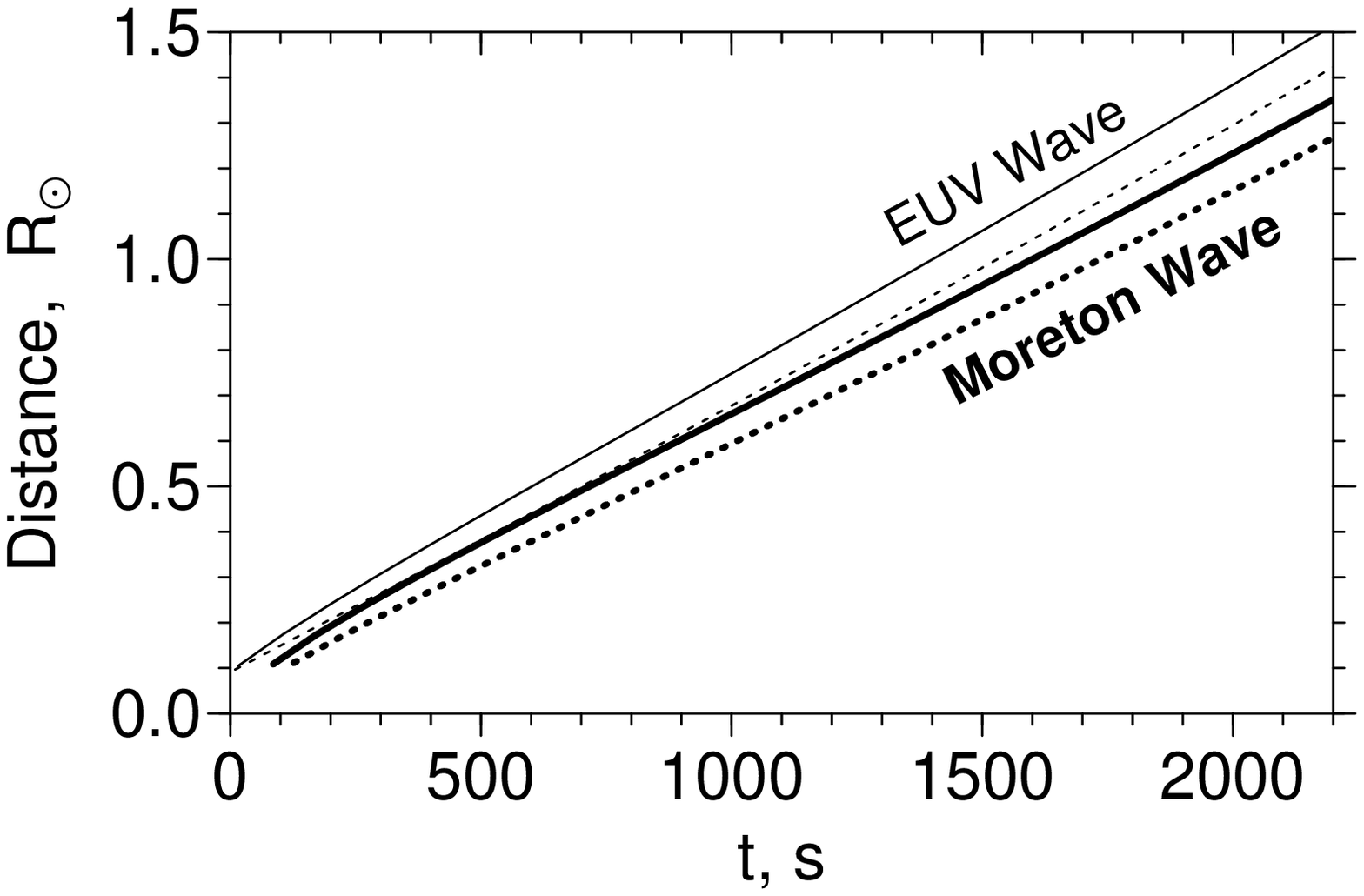}
                \includegraphics[width=0.5\textwidth,clip=]{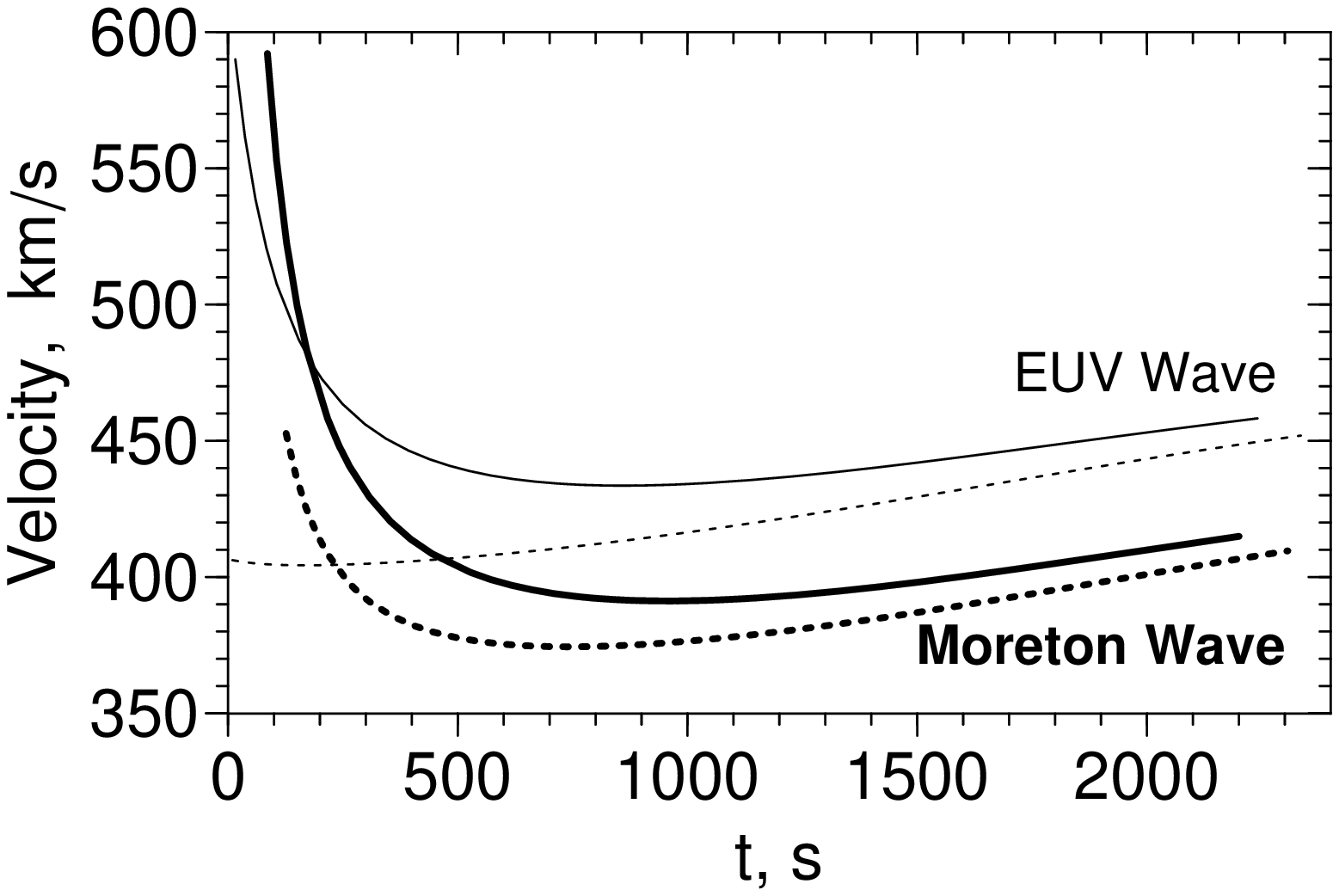}
             }
    \vspace{-0.29\textwidth}
          \centerline{\large
      \hspace{0.125 \textwidth}  {a}
      \hspace{0.47 \textwidth} {b}
         \hfill}
   \vspace{0.255\textwidth}

\caption{The kinematic plots of a Moreton wave (thick) and an EUV
wave (thin), which are produced by a single coronal wave in the
linear (dotted) and nonlinear (solid) consideration.}
  \label{F-EUV_Moreton_wave_speeds}
  \end{figure}

\begin{figure} 
  \centerline{\includegraphics[width=0.5\textwidth]{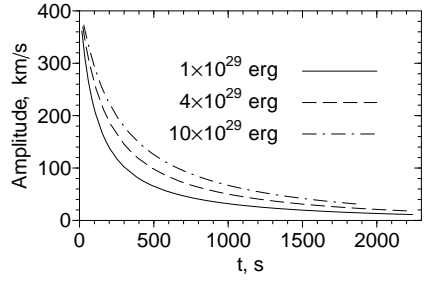} }
\caption{The time evolution of the EUV wave amplitude for different
energies $\tilde{\varepsilon}$ of the wave source.}
  \label{F-EUV_wave_amplitudes}
 \end{figure}

\begin{figure} 
  \centerline{\includegraphics[width=0.5\textwidth]{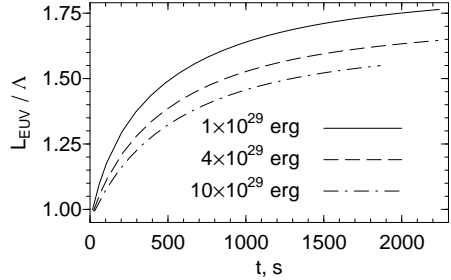} }
\caption{The time dependence of the EUV wave length relative to
its initial size for different energies $\tilde{\varepsilon}$.}
  \label{F-EUV_wavelen}
 \end{figure}

Figures~\ref{F-wave_prop}--\ref{F-EUV_decay} present the results of
our modeling. Figure~\ref{F-wave_prop} illustrates a 3D image of the
shock wave front and the respective 2D section including ray
trajectories. The rays go out from the initial surface of size
$\Lambda$. The wave front gets inclined over the solar surface, with
its inclination increasing in time.

Figure~\ref{F-wave_speed_vs_energy} shows the distance--time plots
of EUV waves (a) as well as the time plots of their velocities (b)
along the surface 60~Mm (relative to the solar surface). The curves
are given for different values of the wave source energy
$\tilde\varepsilon$ or different values of the initial shock wave
length, as follows from $\Lambda=(\tilde\varepsilon/\rho_\ast
a_\ast^2)^{1/3}$. The EUV wave velocity decreases appreciably due to
the nonlinear damping of the coronal wave amplitude. After the
amplitude has substantially decreased, the shock wave propagates as
a linear one. Having reached its minimum, the wave velocity slightly
increases. This is due to the shock front inclination over the solar
surface that becomes more and more significant with time and
associated with the waveguide properties of the lower quiet Sun's
corona. The larger the wave front inclination, the higher the
Moreton and EUV wave velocity. Note that calculated Moreton and EUV
wave acceleration must be difficult to observe since it occurs when
the wave amplitude becomes low (see
Figure~\ref{F-EUV_wave_amplitudes}).

Figure~\ref{F-EUV_Moreton_wave_speeds} shows positions (a) counted
along the surface and velocities (b) of a Moreton wave (thick
lines) and an EUV wave (thin lines), which are produced by a
single coronal wave. The dotted lines correspond to a linear
coronal wave and the solid lines are for a shock one. The linear
Moreton wave velocity is lower because the Alfv{\'e}n speed at the
corona base is smaller than that at a height of $60$ Mm. Note that
a Moreton wave decelerates even in the linear case. This fact is
associated with the initial wave source position at a height of
$80\;Mm$ above the photosphere. The EUV wave in the linear case
does not decelerate since the wave source is located roughly at
the same height that the EUV wave is (a little bit higher the
dashed line in Figure~\ref{F-wave_prop}).

The increase in the propagation velocity of the wave front with
height results in a front inclination relative to the solar surface.
So one can observe the wave signatures corresponding to the
different heights (\textit{e.g.}, EUV waves and Moreton waves) to be
shifted. Increase of the front inclination with time determines a
value of this shift. Besides, its time evolution is also determined
by the observer position and the sight angle. Such a consideration
demonstrates the possibility to explain the offset between Moreton
and EUV waves as well as the HeI-H$\alpha$ offset observed by
\inlinecite{Vrsnak02b} since waves in the He I 10830 $\mathrm \AA$
line are similar morphologically to EUV ones.

Figure~\ref{F-EUV_wave_amplitudes} presents time dependence of the
EUV wave amplitude for different values of the wave source energy
$\tilde\varepsilon$. A disturbance having higher $\tilde\varepsilon$
and greater length decays more slowly as follows
from~(\ref{E-damping_laws}) and as is seen in
Figure~\ref{F-EUV_wave_amplitudes}. The Moreton wave amplitude
varies in a similar manner, but it has smaller values.

Another effect associated with the nonlinearity of an EUV wave is
the increase in its duration $T_{sh}$  and, respectively, length
$L=aT_{sh}$ (also referred to as the wave profile broadening). In
the linear case, the wave duration is constant (under the assumption
of a steady-state medium). Figure~\ref{F-EUV_wavelen} gives the time
evolution of the ratio of the EUV wave length to its initial size
$\Lambda$. If initial amplitudes of disturbances are equal, the
relative extension will be faster for disturbances having the lower
initial energy (and the shorter initial length).

With respect to the damping of shock waves having the same initial
lengths, amplitude decrease is faster for a wave with a higher
amplitude. Therefore, shock waves with different initial
amplitudes decay to the same level after approximately equal
duration (Figure~\ref{F-EUV_decay}, solid curves). For comparison,
we also plot the amplitude curves of linear disturbances, which
hold an initial ratio of the wave amplitudes throughout
propagation. The mentioned property of nonlinear waves (in
Figure~\ref{F-EUV_decay}) allows us to be not precise about the
value of the initial wave amplitude. Therefore, we can use the
value that follows from the strong point-like explosion theory.

\section{Discussion and Conclusion}
\label{S-conclusion}

We have modeled the propagation of shock-associated EUV waves and
Moreton waves, using the nonlinear geometrical acoustics method.
This method takes into account characteristic properties of
nonlinear waves: \textit {i})~dependence of the wave velocity on its
amplitude, \textit {ii})~wave energy dissipation in the shock front,
and \textit {iii})~wave duration increase with time. The method
allows one to calculate the nonlinear evolution of the shock wave
and its propagation pattern. However, the generally accepted variant
of this approach includes nonlinearity effects only for describing
the amplitude and duration of a wave, but not for the wave velocity
value. So, using such a non-self-consistent approach results in the
loss of an important effect concerning wave kinematics.

We have applied another approach, appending an additional term to
the ray equations. This has allowed the finite wave amplitude to be
taken into account. We have solved self-consistently the modified
ray equations and the equations describing the wave amplitude and
duration evolution along a ray tube. It is this approach that has
been developed in this paper to analyze coronal shock wave
propagation along the solar surface.

One of the results of our analysis is the deceleration of EUV and
Moreton waves at the initial stage of propagation. Since we use a
spherically symmetric and isothermal model of the solar corona,
deceleration is a direct consequence of their nonlinearity. Thus,
EUV and Moreton waves having a sufficient amplitude (and therefore
being observable) have to decelerate in the quiet Sun's regions
where average plasma parameters are constant along the solar
surface. Note that the large-scale waves under study are registered
just in these regions. Calculated wave deceleration is supported by
the EUV and Moreton wave observations analyzed by
\citeauthor{Warmuth01} (\citeyear{Warmuth01,Warmuth04a}). Also we
notice here that EUV waves not associated with a coronal MHD wave
show only slight or no deceleration (see, \textit {e.g.},
\opencite{WillsDavey09}).

The simple corona model also let us find other features of wave
kinematics, \textit {e.g.}, \textit {i}) the wave source height has
effect on the initial portion of the velocity plot, and \textit
{ii}) the rate of wave deceleration and damping becomes lower as the
wave source energy (or the wave length, respectively) grows. All
these findings are applied for modeling EUV wave propagation in the
17 January 2010 event in an accompanying paper \cite{Grechnev10III}.

Modeling waves in the linear approximation does not reveal their
deceleration. On the contrary, linear waves undergo only small
acceleration caused by a slightly increasing inclination of the
coronal wave front over the solar surface. This effect was first
discovered by \inlinecite{Uchida68}. However, because of the error
that he made in the expression for the barometric distribution of
the coronal plasma density (the scale height was halved), the wave
front inclination over the solar surface was very large. As a
result, linear waves underwent considerable acceleration in
Uchida's original model.

Another important result of our modeling is the duration (and
length) increase of Moreton and EUV waves. This effect is also
confirmed by observations (see, \textit{e.g.}, \opencite{Warmuth01};
\opencite{Veronig10}). In contrast, a linear disturbance keeps its
duration unchanged in a steady-state medium. Note that in the linear
approximation, the wave amplitude and duration also vary due to the
viscosity, the thermal conductivity and the finite plasma
conductivity, however, these effects are negligible against
nonlinear factors.

To summarize, we believe that wave deceleration and its duration
increase, both being the attributes of shock wave evolution, point
out the crucial role of nonlinearity in the behavior of EUV and
Moreton waves (at least, it concerns some of them).

In conclusion, we will briefly discuss the method limitations for
solving the shock wave propagation problem. The main limitation is
associated with laws~(\ref{E-damping_laws}) of a shock wave
damping, which are derived by using the relations for simple flat
MHD waves in a homogeneous medium. So, we have to meet two
requirements. First, the shock wave length should be smaller than
the radius of curvature of the wave front and the smallest medium
variation scale. The fulfilment of these conditions also ensures
validity of the linear ray approximation (\ref{E-ray_eq_system}),
which involves actually even less limitations. The smallest
variation scale in our modeling is that of plasma density. So, we
are aware that our computation lies at the boundary of
applicability of nonlinear geometrical acoustics since
characteristic shock wave length $\Lambda$ and density scale
$\rho/\left\vert\nabla\rho\right\vert$ are of the same order of
magnitude.

Second, the nonlinear factor $U_{sh}/a$ should be small. Under this
condition, damping laws (\ref{E-damping_laws}) are derived and this
very condition ensures a correct calculation of the terms involving
$U_{sh}$ in the ray equations system (\ref{E-nonlin_ray_eq_system}).
With respect to the limitations of laws (\ref{E-damping_laws}), the
point-like atmospheric explosion theory \cite{Kestenboim74} suggests
that these laws describe satisfactorily spherical shock wave
propagation up to $U_{sh}/a\leq 1$. When we choose the initial value
$U_{sh}=a$ in Section {\ref{S-modeling}}, we do not therefore go
beyond the scope of the application of relations
(\ref{E-damping_laws}). However, ray equations
(\ref{E-nonlin_ray_eq_system}) and equations
(\ref{E-eq_system_for_calc amp}) at $U_{sh}/a\approx1$ are able to
yield an error in calculations. Nevertheless, this error is
insignificant due to the nearly spherical shape of the wave front at
the initial phase of propagation and then it disappears owing to
rapid decrease in $U_{sh}$.

\begin{acks}

We thank Dr.~V.V.~Grechnev for useful discussions and valuable
help in preparing the paper. We also thank the anonymous referee
and editors for careful reading of the manuscript as well as
helpful comments and suggestions. A.A. is very grateful to the
scientific organizing committee of the CESRA2010 Workshop for
financial support.

The research was supported by the Russian Foundation of Basic
Research (Grant No. 10--02--09366) and Siberian Branch of the
Russian Academy of Sciences (Lavrentyev Grant 2010--2011).

\end{acks}

\end{article}
\end{document}